\documentclass[prd,nofootinbib]{revtex4}
\usepackage{epsfig}
\usepackage{graphicx}
\usepackage{dcolumn}
\usepackage{amsmath}
\usepackage{subfig}
\usepackage{latexsym}
\usepackage{hyperref}
\graphicspath{ {./Graphs/} }
\makeatletter
\newcommand*{\rom}[1]{\expandafter\@slowromancap\romannumeral #1@}
\makeatother

\begin{document}
\title{Phase transitions in $D$-dimensional Gauss-Bonnet-Born-Infeld $AdS$
black holes }  

\author{Neeraj Kumar$^{a}$\footnote{e-mail:
neerajkumar@bose.res.in} Sunandan Gangopadhyay$^{a}$\footnote{e-mail: sunandan.gangopadhyay@gmail.com, sunandan.gangopadhyay@bose.res.in}}
\affiliation{$^a$Department of Theoretical Sciences, S.N.Bose National Centre for Basic Sciences,\\ JD Block, Sector III, Salt Lake, Kolkata, 700106, India\\}
\maketitle
\section*{Abstract}
\noindent In this paper, we have investigated the phase transition in black holes when Gauss-Bonnet corrections to the spacetime curvature and Born-Infeld extension in stress-energy tensor of electromagnetic field are considered in a negative cosmological constant background. It is evident that the black hole undergoes a phase transition as the specific heat capacity at constant potential shows discontinuities. Further, 
the computation of the free energy of the black hole, the Ehrenfest scheme and the Ruppeiner state space geometry analysis are carried out to establish the second order nature of this phase transition. The effect of non-linearity arising from Born-Infeld electrodynamics is also evident from our analysis. Our investigations are done in general $D$-spacetime dimensions with $D>4 $, and specific computations have been carried out in $D= 5,~6,~7$ spacetime dimensions.



\section{Introduction} 
\noindent Singularities are inevitable in Einstein's gravity and the object called black hole has properties which are not fully explained by classical gravity. Hawking \cite{hawking} and Bekenstein \cite{beke} considered these objects as thermodynamic systems and associated with it thermodynamic properties like entropy and temperature. The study of black hole thermodynamics has received a new impetus recently because it is believed that it may shed light on quantum gravity.\

\noindent In seeking modifications to Einstein's gravity in bottom-up approach, the very first idea was to see higher order effects of curvature in the action. This modification was brought in \cite{love} where general spacetime Lagrangian density has higher powers of curvature and is known as Lovelock gravity. These modifications are incorporated in such a way that the solutions resulting from these are consistent and are free from any ghost terms, hence arbitrary terms in the action are not allowed. 
Expansion of the Lovelock Lagrangian contains the Gauss-Bonnet term. 
It is believed that this simple modification may lead to interesting insights into the true nature of gravity.

\noindent Non-linear electrodynamics was first developed by Max Born and Leopold Infeld \cite{max} in order to resolve the self-energy problem of a point charge. However, all such non-linear theories got soon replaced by the profound quantum description of Maxwell's theory in the form of quantum electrodynamics. Later, Born-Infeld theory reappeared as a low energy limit of string theory in \cite{bstring1}-\cite{bstring2}. Since then an extensive study of black holes with these higher derivatives in gauge field has been done \cite{biblack1}-\cite{biblack3}. \

\noindent Thermodynamics of the black holes in Gauss-Bonnet gravity have been investigated in great detail in \cite{g1}-\citep{g6} and the list is not exhaustive. These black holes have been studied with different horizon topologies, with and without charge, with non-linear electrodynamics, with different cosmological constant signature, etc. Heat capacity of Gauss-Bonnet $AdS$ black holes have been computed in \cite{met} but the classification of phase-transitions into first or second order was not done. Our goal is to check the phase structure of these black holes in AdS background with Born-Infeld electrodynamics and analyse it using standard tools of thermodynamics. More precisely, our main aim is to study the thermodynamics of these black holes and figure out the order of the phase-transitions occurring in such black holes.

\noindent In this paper, we wish to study the black hole solutions obtained in Gauss-Bonnet gravity incorporating Born-Infeld theory in the action with negative cosmological constant. All analysis is done in $D$ spacetime dimensions. Thermodynamic quantities are calculated and singularities in heat capacity are registered for the fixed Gauss-Bonnet parameter, Born-Infeld parameter and charge. These divergences are shown in graphs when the heat capacity is plotted with the horizon radius for $D=5,6,7$. As we shall show later that this is a clear indication of a second order phase transition of the black hole system under consideration with specific values of the parameters. Along the way, the effects of non-linearity introduced through Born-Infeld electrodynamics on temperature as well as on phase transition points are also analysed. Ehrenfest scheme and the Ruppeiner state space geometry analysis \cite{ru1}-\cite{ru2} are trusted techniques in standard thermodynamics and are employed to study black hole thermodynamics as well \cite{n1}-\cite{r2}. We wish to employ these for our system and study the order of the phase transitions in the black holes. A parameter called Prigogine-Defay ratio \cite{pd} which determines the deviations from the second order nature of phase transition is also calculated using Ehrenfest equations. All results for Einstein-Born-Infeld black holes in \cite{n1} are recovered when the Gauss-Bonnet parameter is put to zero. Further, motivations to study these black hole phase transitions in AdS space comes from AdS/CFT correspondence \cite{ads1}-\cite{ads3}. Confinement problem, superconductivity and other strongly correlated systems are in the domain of this duality and has led to deep insights to these phenomena.
Along the way, we shall also talk about the possibility of treating the cosmological constant, Born-Infeld parameter and Gauss-Bonnet parameter as thermodynamical variables and derive a Smarr relation using scaling argument and first law of thermodynamics in $D$-spacetime dimensions.

\noindent The structure of this paper is as follows. In section \rom{2} we shall be calculating the thermodynamic properties of the black holes considered in this paper and also calculate their heat capacity. We plot the heat capacity with the horizon radius to study its nature. Section \rom{3} is devoted to carrying out Ehrenfest scheme analysis at the points of phase transitions in order to understand the order of phase transition. Section \rom{4} contains the Ruppeiner state space geometry analysis of the singularities. We conclude by summarizing our results in section \rom{5}. 


   
\section{Thermodynamics of Gauss-Bonnet-Born-Infeld black holes}
\noindent In this section, we introduce the black hole spacetime we are interested in and study its thermodynamic properties. We consider the Gauss-Bonnet black hole spacetime with Born-Infeld electromagnetic charge in AdS background.
The action of Gauss-Bonnet gravity with Born-Infeld electromagnetic field reads 
\begin{eqnarray}
I=\dfrac{1}{16\pi}\int d^Dx\sqrt{-g}[R-2\Lambda +\alpha L_{GB}+L(F)]
\label{a}
\end{eqnarray}
where $G=c=1$, the Gauss-Bonnet Lagrangian density $L_{GB}=R^2-4R_{\gamma\delta}R^{\gamma\delta}+R_{\gamma\delta\lambda\sigma}R^{\gamma\delta\lambda\sigma}$,
the Born-Infeld term $L(F)=4b^2\left( 1-\sqrt{1+\dfrac{F^{\mu\nu}F_{\mu\nu}}{2b^2}}\right)$,
$\alpha$ is the Gauss-Bonnet parameter, $b$ is the Born-Infeld parameter, $\Lambda=-\dfrac{(D-1)(D-2)}{2l^2}$ with $l$ being AdS radius and $D$ is the spacetime dimensions greater than $4$.\

\noindent The solution following from the above theory reads 
\begin{eqnarray}
ds^2=-f(r)dt^2+\dfrac{1}{f(r)}dr^2+r^2d\Omega^2_{D-2}
\label{d}
\end{eqnarray}

\noindent where the metric coefficient $f(r)$ is given by \cite{met}
\begin{eqnarray}
f(r)=1+\dfrac{r^2}{2\alpha'}(1-\sqrt{g(r)})
\label{e}
\end{eqnarray}
with $g(r)$ given by
\begin{eqnarray}
g(r)=1-\dfrac{4\alpha'}{l^2} + \dfrac{4\alpha' m}{r^{D-1}}-\dfrac{16\alpha' b^2}{(D-1)(D-2)}\left(1-\sqrt{1+\dfrac{(D-2)(D-3)q^2}{2b^2r^{2D-4}}}\right)\nonumber\\
-\dfrac{8(D-2)\alpha' q^2}{(D-1)r^{2D-4}}{}_2F_1\left[\dfrac{D-3}{2D-4},\dfrac{1}{2},\dfrac{3D-7}{2D-4},-\dfrac{(D-2)(D-3)q^2}{2b^2r^{2D-4}}\right].
\label{f}
\end{eqnarray}
Note that the actual black hole parameters such as the charge $Q$, mass $M$  and Gauss-Bonnet coefficent $\alpha$ are connected to $q$, $m$ and $\alpha'$ as
\begin{eqnarray}
M &=& \dfrac{(D-2)\omega}{16\pi}m,~~~~\alpha=\dfrac{\alpha'}{(D-3)(D-4)}~, \nonumber\\ 
Q&=&\sqrt{2(D-2)(D-3)}\dfrac{\omega}{8\pi}q,~~~~~~\omega=\dfrac{2\pi^{\dfrac{D-1}{2}}}{\Gamma\dfrac{(D-1)}{2}}~.
\label{g}
\end{eqnarray}
The black hole mass can be written in terms of the horizon radius ($r_+$) from the condition $f(r_+)=0$ and reads
\begin{eqnarray}
M=\dfrac{(D-2)\omega\alpha'}{16\pi}r_+^{D-5}+\dfrac{(D-2)\omega}{16\pi}r_+^{D-3}+\dfrac{(D-2)\omega}{16\pi l^2}r_+^{D-1}+\dfrac{b^2\omega}{4\pi(D-1)}r_+^{D-1}\left(1-\sqrt{1+\dfrac{16\pi^2Q^2}{b^2\omega^2r_+^{2D-4}}}\right)\nonumber\\
+\dfrac{4(D-2)\pi Q^2}{(D-1)(D-3)\omega r_+^{D-3}}{}_2F_1\left[\dfrac{D-3}{2D-4},\dfrac{1}{2},\dfrac{3D-7}{2D-4},-\dfrac{16\pi^2Q^2}{b^2\omega^2r_+^{2D-4}}\right].
\label{h}
\end{eqnarray}
The Hawking temperature of the black hole spacetime can be calculated using eq.(\ref{e}) as
\begin{eqnarray}
T &=&\dfrac{1}{4\pi}\left(\dfrac{df(r)}{dr}\right)_{r_+} \nonumber\\
&=&\dfrac{1}{4\pi r_+(r^2_++2\alpha')}\left[\dfrac{(D-1)r_+^4}{l^2}+(D-3)r^2_++(D-5)\alpha'+\dfrac{4b^2r_+^4}{D-2}\left(1-\sqrt{1+\dfrac{16\pi^2 Q^2}{b^2\omega^2 r_+^{2D-4}}}\right)\right]~
\label{i}
\end{eqnarray}
where $l$ shall be fixed to unity for rest of the calculations unless mentioned otherwise.\

\begin{figure}[!ht]
		\subfloat[D=5\label{f1}]{{\includegraphics[scale=0.50]{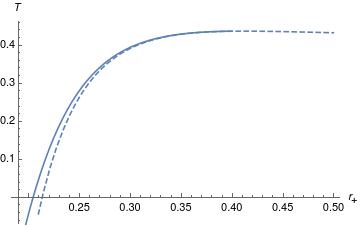}}}
		\qquad
		~~~~~~~~~~
		\subfloat[D=6\label{f2}]{{\includegraphics[scale=0.50]{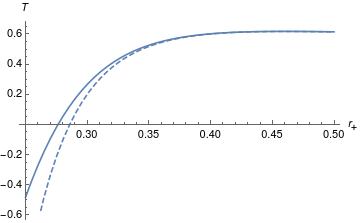}}}
		\qquad
		~~~~~~~~~~
		\subfloat[D=7\label{f3}]{{\includegraphics[scale=0.50]{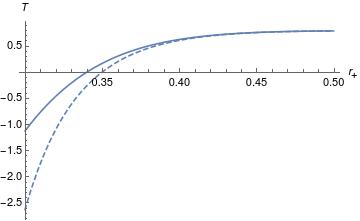}}}
		\qquad
		\caption{Temperature vs. horizon radius}{Solid line(Q=0.13, b=10, $\alpha'$=0.01)}\\{Dashed line(Q=0.13, b$\rightarrow\infty$, $\alpha'$=0.01)}
		\label{fig1}
\end{figure}

\noindent We now use the relation
\begin{eqnarray}
dM=TdS+\Phi dQ
\end{eqnarray}
which is the first law of black hole thermodynamics. This form is analogous to the first
law of thermodynamics
\begin{eqnarray}
dU=TdS-PdV
\end{eqnarray}
with the identification of the pressure $P$ to the negative of the electrostatic potential $\Phi$,
the volume to the charge $Q$ and the internal energy $U$ to the mass of the black hole $M$.
From this relation, the black hole entropy $S$ can be obtained as 
\begin{eqnarray}
S=\int_0^{r_{+}} \frac{1}{T} \left(\frac{\partial M}{\partial r_+}\right)_Q dr_+=\dfrac{\omega}{4}r_+^{D-2}\left[1+\dfrac{D-2}{D-4}\dfrac{2\alpha'}{r_+^2}\right]~.
\label{j}
\end{eqnarray}
\noindent In principle, eq.(\ref{j}) allows the Hawking temperature ($T$) of the black hole to be written in terms of the entropy $S$.
\noindent In Fig.\ref{fig1}, we plot the Hawking temperature of the black hole with the horizon radius and observe that there is no discontinuity in these graphs hence there is no first order phase transition. The Maxwell limit is also displayed for these black holes. It can be seen that the black holes with Born-Infeld electrodynamics have higher temperature than with Maxwell electrodynamics. This effect is prominent for smaller black holes and fades away as the horizon radius increases.

\noindent 
Further, we analyse the system and calculate the heat capacity at constant potential. This we do to find whether there is a chance of higher order phase transitions in this black hole spacetime.\

\noindent The potential in $D$ dimensions for the black hole spacetime can be calculated from the first law and eq.(\ref{h}) to be
\begin{eqnarray}
\Phi=\left(\dfrac{\partial M}{\partial Q}\right)=\dfrac{4\pi Q}{\omega(D-3)r_+^{D-3}}{}_2F_1\left[\dfrac{D-3}{2D-4},\dfrac{1}{2},\dfrac{3D-7}{2D-4},-\dfrac{16\pi^2Q^2}{b^2\omega^2r_+^{2D-4}}\right]~.
\end{eqnarray} 

\noindent Considering the temperature of the black hole to be a function of entropy and charge  ($T\equiv T(S,Q)$), we have
\begin{eqnarray}
\left(\dfrac{\partial T}{\partial S}\right)_{\Phi}=\left(\dfrac{\partial T}{\partial S}\right)_{Q}-\left(\dfrac{\partial T}{\partial Q}\right)_{S}\left(\dfrac{\partial \Phi}{\partial S}\right)_{Q}\left(\dfrac{\partial Q}{\partial \Phi}\right)_{S}~.
\label{m}
\end{eqnarray}
Now using the thermodynamic identity
\begin{eqnarray}
\left(\dfrac{\partial Q}{\partial S}\right)_{\Phi}\left(\dfrac{\partial S}{\partial \Phi}\right)_{Q}\left(\dfrac{\partial \Phi}{\partial Q}\right)_{S}=-1
\label{n}
\end{eqnarray}
along with eq.(\ref{m}), we obtain the heat capacity at constant potential to be
\begin{eqnarray}
C_{\Phi} &=&T\left(\dfrac{\partial S}{\partial T}\right)_{\Phi}\nonumber\\
&=&\dfrac{T\left(\dfrac{\partial \Phi}{\partial Q}\right)_{r_+}\left(\dfrac{\partial S}{\partial r_+}\right)_{Q}}{\left(\dfrac{\partial \Phi}{\partial Q}\right)_{r_+}\left(\dfrac{\partial T}{\partial r_+}\right)_Q-\left(\dfrac{\partial T}{\partial Q}\right)_{r_+}\left(\dfrac{\partial \Phi}{\partial r_+}\right)_Q}~.
\label{o}
\end{eqnarray}
The partial derivatives involved in the above equation when calculated read 
\begin{eqnarray}
\left(\dfrac{\partial \Phi}{\partial r_+}\right)_{Q}=-\dfrac{4\pi Q}{\omega r_+^{D-2}}\left(1+\dfrac{16\pi^2Q^2}{b^2\omega^2r_+^{2D-4}}\right)^{-1/2}
\label{p}
\end{eqnarray}
\begin{eqnarray}
\left(\dfrac{\partial T}{\partial Q}\right)_{r_+}=-\dfrac{1}{r_+^2+2\alpha'}\dfrac{16\pi Q}{\omega^2(D-2)r_+^{2D-7}}\left(1+\dfrac{16\pi^2Q^2}{b^2\omega^2r_+^{2D-4}}\right)^{-1/2}
\label{q}
\end{eqnarray}
\begin{eqnarray}
\left(\dfrac{\partial S}{\partial r_+}\right)_{Q}\equiv\dfrac{dS}{dr_+}=\dfrac{\omega}{4}(D-2)r_+^{D-3}\left(1+\dfrac{2\alpha'}{r_+^2}\right)
\label{r}
\end{eqnarray}
\begin{eqnarray}
\left(\dfrac{\partial \Phi}{\partial Q}\right)_{r_+}=\dfrac{4\pi}{(D-2)\omega r_+^{D-3}}\left[\left(1+\dfrac{16\pi^2Q^2}{b^2\omega^2r_+^{2D-4}}\right)^{-1/2}+\dfrac{1}{D-3}{}_2F_1\left[\dfrac{D-3}{2D-4},\dfrac{1}{2},\dfrac{3D-7}{2D-4},-\dfrac{16\pi^2Q^2}{b^2\omega^2r_+^{2D-4}}\right]\right]
\label{s}
\end{eqnarray}
\begin{eqnarray}
\left(\dfrac{\partial T}{\partial r_+}\right)_Q=\dfrac{1}{4\pi r_+(r_+^2+2\alpha')}\left[4(D-1)r_+^3+2(D-3)r_++\dfrac{16b^2r^3}{(D-2)}\left(1-\sqrt{1+\dfrac{16\pi^2Q^2}{b^2\omega^2r_+^{2D-4}}}\right)\right. \nonumber\\
\left.+\dfrac{64\pi^2Q^2}{\omega^2r_+^{2D-7}}\left(1+\dfrac{16\pi^2Q^2}{b^2\omega^2r_+^{2D-4}}\right)^{-1/2}\right]\nonumber\\
-\dfrac{3r_+^2+2\alpha'}{4\pi r_+^2(r_+^2+2\alpha')^2}\left[(D-1)r_+^4+(D-3)r^2_++(D-5)\alpha'+\dfrac{4b^2r_+^4}{D-2}\left(1-\sqrt{1+\dfrac{16\pi^2 Q^2}{b^2\omega^2 r_+^{2D-4}}}\right)\right]~.
\label{t}
\end{eqnarray}
The form of heat capacity is a large expression to write, hence we study its behaviour through plots. We plot $C_{\Phi}$ with horizon radius $(r_+)$ for fixed values of the Gauss-Bonnet parameter $(\alpha')$, Born-Infeld parameter $(b)$ and charge $(Q)$ for $D=5,~6,~7$.

\vspace*{3mm}

\begin{figure}[!ht]
		\subfloat[D=5\label{f1}]{{\includegraphics[scale=0.50]{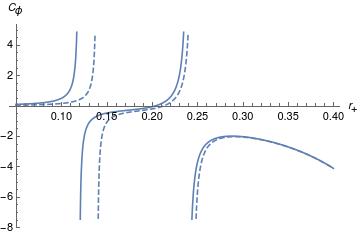}}}
		\qquad
		~~~~~~~~~~
		\subfloat[D=5\label{f1}]{{\includegraphics[scale=0.50]{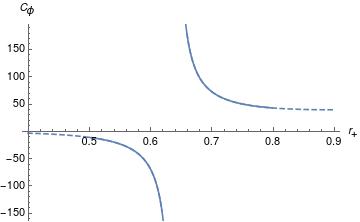}}}
		\qquad
		~~~~~~~~~~
		\subfloat[D=6\label{f2}]{{\includegraphics[scale=0.50]{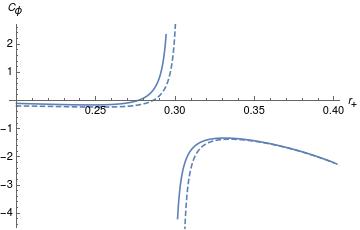}}}
		\qquad
		~~~~~~~~~~
		\subfloat[D=6\label{f1}]{{\includegraphics[scale=0.50]{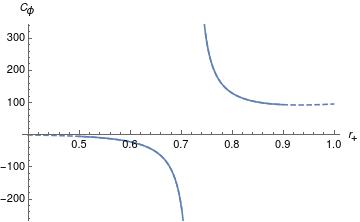}}}
		\qquad
		~~~~~~~~~~
		\subfloat[D=7\label{f1}]{{\includegraphics[scale=0.50]{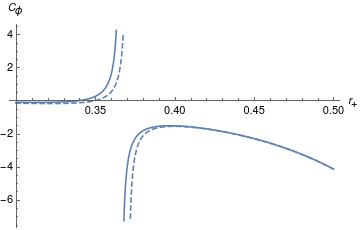}}}
		\qquad
		~~~~~~~~~~
		\subfloat[D=7\label{f3}]{{\includegraphics[scale=0.50]{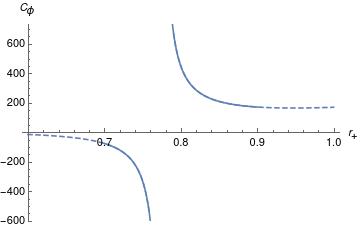}}}
		\qquad
		\caption{Heat capacity vs. horizon radius}{Solid line(Q=0.13, b=10, $\alpha'$=0.01)}\\{Dashed line(Q=0.13, b$\rightarrow\infty$, $\alpha'$=0.01)}
\end{figure}

\noindent The plots of heat capacity have multiple points of discontinuities and give indications of a second order phase transition. We calculate the exact points of discontinuities of heat capacity. We also calculate the horizon radius of extremal black hole ($r^{(e)}_+$) for the same fixed parameters for each dimension in order to check the physical validity of the singularities in heat capacity. The following Table shows extremal black hole radii ($r^{(e)}_+$) and singularities in heat capacity at radii $r_{+0},~r_{+1},$ and $r_{+2}$.
\begin{center}
\begin{tabular}{ |c|c|c|c| }
\hline
 	   & D=5      & D=6       & D=7        \\  	   
\hline 
 $r^{(e)}_+$		&0.204957&0.276397&0.340496\\
\hline
 $r_{+0}$	& 0.119316		&--	 &  --\\
 $r_{+1}$   & 0.23969 & 0.298154 & 0.365921  \\ 
 $r_{+2}$   & 0.639584  & 0.725395   & 0.774541  \\
 \hline
\end{tabular}
\captionof{table}{Extremal radii and values of horizon radius where $C_{\Phi}$ diverge.\\For solid lines (Q=0.13, b=10, $\alpha'$=0.01)}
\label{tab1}
\end{center}
Table \ref{tab1} shows that the black hole in $D=5$ spacetime dimensions has one unphysical singularity ($r_{+0}$) in the heat capacity as the point is below the extremal radius whereas the other two at points $r_{+1}$ and $r_{+2}$ are actual phase transition points. For $D=6,7$ there are two singularities each which are actual phase transition points as both the points lie above extremal radius values for respective dimensions. These plots also reveal that for a particular spacetime dimension there are three phases of the black hole, namely phase I ($r^{(e)}_{+} < r_+ < r_{+1}$), phase II ($r_{+1} < r_+ < r_{+2}$), and phase III ($r_{+} > r_{+2}$). The dashed curves in the plots depict the behaviour of these black holes in the Maxwell limit. It is apparent that the Born-Infeld non-linearity shifts the phase transition points to lower value of the horizon radius and this effect is significant at smaller horizon radius.\

\noindent In the following section, we shall show that the system follows Ehrenfest second order phase transition equations. We shall also carry out the Ruppeiner's state space geometry analysis to confirm the second order phase transition exhibited by this black hole spacetime.\

\noindent We would now like to discuss about the possibility of interpreting the cosmological constant $\Lambda$ as a thermodynamic pressure $p$ and also treating the Born-Infeld parameter $b$ and the Gauss-Bonnet parameter $\alpha$ as new thermodynamic variables, as it has been proposed recently in \cite{dolan}-\cite{maan}. The argument put forward in these studies was that since $\Lambda$, $b$ and $\alpha$ are dimensionful quantities, the corresponding terms would definitely appear in the Smarr formula. To write down the Smarr formula, we first note that the first law of black hole thermodynamics takes the form 
\begin{eqnarray}
dM=TdS+vdp+\Phi dQ+Bdb+\Omega d\alpha
\label{ii}
\end{eqnarray}
where $B$ and $\Omega$ are the conjugate variables to $b$ and $\alpha$ respectively defined by
\begin{eqnarray}
B=\dfrac{\partial M}{\partial b}~~~~~and~~~~~~\Omega=\dfrac{\partial M}{\partial \alpha}
\end{eqnarray}

\noindent Considering the black hole mass $M = M (S, Q, p, b, \alpha)$ and performing dimensional analysis,
we find that $[M]=L^{D-3},~~[S]=L^{D-2},~~[p]=L^{-2},~~[Q]=L^{D-3}$,~~$[b]=L^{-1}$ and $[\alpha]=L^{2}$. 
Using these along with Euler's theorem\footnote[1]{Given a function $g(x,y)$ satisfying $g(\alpha^m x,\alpha^n y)=\alpha^rg(x,y)$, we have $rg(x,y)=mx\left(\dfrac{\partial g}{\partial x}\right)+ny\left(\dfrac{\partial g}{\partial y}\right)$.}, we obtain
\begin{eqnarray}
(D-3)M=(D-2)S\left(\dfrac{\partial M}{\partial S}\right)-2p\left(\dfrac{\partial M}{\partial p}\right)+(D-3)Q\left(\dfrac{\partial M}{\partial Q}\right)-b\left(\dfrac{\partial M}{\partial b}\right)+2\alpha\left(\dfrac{\partial M}{\partial \alpha}\right)~.
\label{sma}
\end{eqnarray}
Now using eq.(\ref{ii}), we get
\begin{eqnarray}
\left(\dfrac{\partial M}{\partial S}\right)=T,~~\left(\dfrac{\partial M}{\partial p}\right)=v,~~\left(\dfrac{\partial M}{\partial Q}\right)=\Phi,~~\left(\dfrac{\partial M}{\partial b}\right)=B,~~\left(\dfrac{\partial M}{\partial \alpha}\right)=\Omega ~.
\end{eqnarray}

\noindent Substituting this in eq.(\ref{sma}) yields the Smarr formula 
\begin{eqnarray}
M=\dfrac{D-2}{D-3}TS-\dfrac{2}{D-3}pv+\Phi Q-\dfrac{1}{D-3}bB+\dfrac{2}{D-3}\alpha \Omega~.
\end{eqnarray}
\noindent It would be interesting to work with the first law of black hole thermodynamics given in eq.(\ref{ii}) and study phase transitions. However, we shall not carry out this investigation in this work and we intend to do it in future.

\noindent Before ending this section, we would like to compute the free energy 
($F=M-TS$) of this black hole. Using eq.(s)(\ref{h},~\ref{i} and \ref{j}), 
this takes the form
\begin{eqnarray}
F=\dfrac{(D-2)\omega r_+^{D-3}}{16\pi}\left(1+r_+^2+\dfrac{\alpha '}{r_+^2}\right)+\dfrac{b^2\omega}{4\pi(D-1)}r_+^{D-1}\left(1-\sqrt{1+\dfrac{16\pi^2Q^2}{b^2\omega^2r_+^{2D-4}}}\right)+\dfrac{4(D-2)\pi Q^2}{(D-1)(D-3)\omega r_+^{D-3}}{}_2F_1\nonumber\\
-\dfrac{\omega r_+^{D-1}}{16\pi(D-4)}\dfrac{\left((D-4)r_+^2+2(D-2)\alpha '\right)}{r_+^2+2\alpha'}\left[(D-1)r_+^4+(D-3)r^2_++(D-5)\alpha'+\dfrac{4b^2r_+^4}{D-2}\left(1-\sqrt{1+\dfrac{16\pi^2 Q^2}{b^2\omega^2 r_+^{2D-4}}}\right)\right]~
\label{free}
\end{eqnarray}
where ${}_2F_1\equiv{}_2F_1\left[\dfrac{D-3}{2D-4},\dfrac{1}{2},\dfrac{3D-7}{2D-4},-\dfrac{16\pi^2Q^2}{b^2\omega^2r_+^{2D-4}}\right]$.\

\noindent The above expression for $D=5$ simplifies to
\begin{eqnarray}
F=\dfrac{3\pi r_+^2}{8}\left(1+r_+^2+\dfrac{\alpha'}{r_+^2}\right)+\dfrac{b^2\pi r_+^4}{8}\left(1-\sqrt{1+\dfrac{4Q^2}{b^2\pi^2 r_+^6}}\right)+\dfrac{3Q^2}{4\pi r_+^2}{}_2F_1\left[\dfrac{1}{3},\dfrac{1}{2},\dfrac{4}{3},-\dfrac{4Q^2}{b^2\pi^2r_+^{6}}\right]\nonumber\\
-\dfrac{\pi r_+^4}{8}\left(\dfrac{r_+^2+6\alpha'}{r_+^2+2\alpha'}\right)\left[4r_+^4+2r_+^2+\dfrac{4b^2r_+^4}{3}\left(1-\sqrt{1+\dfrac{4Q^2}{b^2\pi^2r_+^6}}\right)\right]~.
\end{eqnarray}
\begin{figure}[!ht]
		\subfloat[D=5\label{f1}]{{\includegraphics[scale=0.70]{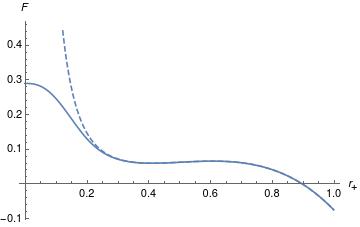}}}
		\qquad
	~~~~
		\caption{Free Energy vs. horizon radius}{Solid line(Q=0.13, b=10, $\alpha'$=0.01)}\\{Dashed line(Q=0.13, b$\rightarrow\infty$, $\alpha'$=0.01)}
		\label{fr}
\end{figure}
Fig.[\ref{fr}] shows the plot of free energy with horizon radius and does not show any cusp. This excludes the possibility of a first order phase transition. In the next section we shall employ the Ehrenfest scheme of analysing phase transitions and establish that the discontinuities shown in the heat capacity of the black hole corresponds to second order phase transitions.



\section{Analysis of phase transition using Ehrenfest scheme}
\noindent Ehrenfest's approach to studying phase transitions is the standard technique in thermodynamics to determine the nature of phase transitions for various thermodynamical systems \cite{e1}, \cite{e2}. It simply says that the order of the phase transition corresponds to the discontinuity in the order of the derivative of Gibb's potential. For the second order phase transition the second derivative encounters discontinuities, however first derivative and Gibb's potential at those points are continuous. These conditions with Maxwell's relations give two equations which have to be satisfied for the second order phase transition.\

\noindent The first and second Ehrenfest equations in thermodynamics are given by
\begin{eqnarray}
\left(\dfrac{\partial P}{\partial T}\right)_S&=&\dfrac{1}{VT}\dfrac{C_{P_2}-C_{P_1}}{\beta_2-\beta_1}=\dfrac{\Delta C_P}{VT\Delta\beta},
\label{u}
\end{eqnarray}
\begin{eqnarray}
\left(\dfrac{\partial P}{\partial T}\right)_V&=&\dfrac{\beta_2-\beta_1}{\kappa_2-\kappa_1}=\dfrac{\Delta\beta}{\Delta\kappa}
\label{v}
\end{eqnarray}
where subscripts $1$ and $2$ denote two distinct phases of the system. Now we use the correspondence between the pressure ($P$) to the negative of the electrostatic potential difference ($\Phi$) and the volume ($V$) to the charge ($Q$) of the black hole. These identifications lead to the following equations 
\begin{eqnarray}
-\left(\dfrac{\partial \Phi}{\partial T}\right)_S&=&\dfrac{1}{QT}\dfrac{C_{{\Phi}_2}-C_{\Phi_1}}{\beta_2-\beta_1}=\dfrac{\Delta C_{\Phi}}{QT\Delta\beta}
\label{w}
\end{eqnarray}
\begin{eqnarray}
-\left(\dfrac{\partial \Phi}{\partial T}\right)_Q&=&\dfrac{\beta_2-\beta_1}{\kappa_2-\kappa_1}=\dfrac{\Delta\beta}{\Delta\kappa}~.
\label{x}
\end{eqnarray}
Note that $\beta$ is the volume expansion coefficient and $\kappa$ is the isothermal compressibility of the system and are defined as
\begin{eqnarray}
\beta=\dfrac{1}{Q}\left(\dfrac{\partial Q}{\partial T}\right)_{\Phi}~,~~~~
\kappa=\dfrac{1}{Q}\left(\dfrac{\partial Q}{\partial \Phi}\right)_{T}~.
\label{y}
\end{eqnarray}
\noindent Now we proceed to check whether the black hole phase transition satisfies the Ehrenfest equations (\ref{w}, \ref{x}). In other words, we investigate the validity of the Ehrenfest equations at the points of discontinuities $r_{+i}$,~$i=1, 2$. Here we denote the critical values of temperature by $T_i$, charge by $Q_i$ and entropy by $S_i$.
\noindent The left hand side of the first Ehrenfest equation (\ref{w}) at the critical point can be written as
\begin{eqnarray}
-\left[\left(\dfrac{\partial \Phi}{\partial T}\right)_S\right]_{S=S_i}\equiv-\left[\left(\dfrac{\partial \Phi}{\partial T}\right)_{r_+}\right]_{r_{+}=r_{+i}}&=&-\left[\left(\dfrac{\partial \Phi}{\partial Q}\right)_{r_+}\right]_{r_+=r_{+i}}\left[\left(\dfrac{\partial Q}{\partial T}\right)_{r_+}\right]_{r_+=r_{+i}} \nonumber\\ 
&=&-\dfrac{\left[\left(\dfrac{\partial \Phi}{\partial Q}\right)_{r_+}\right]_{r_+=r_{+i}}}{\left[\left(\dfrac{\partial T}{\partial Q}\right)_{r_+}\right]_{r_+=r_{+i}}}~.
\label{z}
\end{eqnarray}
Using eq.(s)(\ref{q},~~\ref{s}) we calculate the left hand side of the above relation and obtain
\begin{eqnarray}
-\left[\left(\dfrac{\partial \Phi}{\partial T}\right)_{r_+}\right]_{r_{+}=r_{+i}}=-\dfrac{\omega (r_{+i}^2+2\alpha')r_{+i}^{D-4}}{4Q_i}\left[1+\dfrac{1}{D-3}\sqrt{1+\dfrac{16\pi^2Q_i^2}{b^2\omega^2r_{+i}^{2D-4}}}{}_2F_1\left[\dfrac{D-3}{2D-4},\dfrac{1}{2},\dfrac{3D-7}{2D-4},-\dfrac{16\pi^2Q_i^2}{b^2\omega^2r_{+i}^{2D-4}}\right]\right].
\label{a1}
\end{eqnarray}

\noindent Using eq.(\ref{y}) and the definition of heat capacity $C_{\Phi}= T(\frac{\partial S}{\partial T})_{\Phi}$, we can obtain the right hand side of eq.(\ref{w}) to be 
\begin{eqnarray} 
Q_i\beta=\left[\left(\dfrac{\partial Q}{\partial T}\right)_{\Phi}\right]_{S=S_i}=\left[\left(\dfrac{\partial Q}{\partial S}\right)_{\Phi}\right]_{S=S_i}\left(\dfrac{C_{\Phi}}{T_i}\right)
\label{a2}
\end{eqnarray}
which implies
\begin{eqnarray}
\dfrac{\Delta C_{\Phi}}{T_i Q_i\Delta\beta}=\left[\left(\dfrac{\partial S}{\partial Q}\right)_{\Phi}\right]_{S=S_i}~.
\label{a3}
\end{eqnarray}
\noindent Using the identity (\ref{n}), the above equation can be written in the form
\begin{eqnarray}
\dfrac{\Delta C_{\Phi}}{T_i Q_i\Delta\beta}&=&-\dfrac{
\left[\left(\dfrac{\partial \Phi}{\partial Q}\right)_S\right]_{S=S_i}}{\left[\left(\dfrac{\partial \Phi}{\partial S}\right)_Q\right]_{S=S_i}}\nonumber\\
&\equiv &-\dfrac{
\left[\left(\dfrac{\partial \Phi}{\partial Q}\right)_{r_+}\right]_{r_+=r_{+i}}\left[\dfrac{dS}{dr_+}\right]_{r_+=r_{+i}}}{\left[\left(\dfrac{\partial \Phi}{\partial r_+}\right)_Q\right]_{r_+=r_{+i}}}~.
\label{a4}
\end{eqnarray}
Now using eq.(s) (\ref{p}, \ref{r} , \ref{s}), we calculate the right hand side of the above relation to be 
\begin{eqnarray}
\dfrac{\Delta C_{\Phi}}{T_i Q_i\Delta\beta}=-\dfrac{\omega (r_{+i}^2+2\alpha')r_{+i}^{D-4}}{4Q_i}\left[1+\dfrac{1}{D-3}\sqrt{1+\dfrac{16\pi^2Q_i^2}{b^2\omega^2r_{+i}^{2D-4}}}{}_2F_1\left[\dfrac{D-3}{2D-4},\dfrac{1}{2},\dfrac{3D-7}{2D-4},-\dfrac{16\pi^2Q_i^2}{b^2\omega^2r_{+i}^{2D-4}}\right]\right]~.
\label{a5}
\end{eqnarray}
Eq.(\ref{a1}) and eq.(\ref{a5}) show the validity of the first Ehrenfest's equation for the black hole spacetime under consideration. We now proceed to check the second Ehrenfest equation.
To calculate the left hand side of the second Ehrenfest equation, we use the thermodynamic relation 
\begin{eqnarray}
\left(\dfrac{\partial T}{\partial \Phi}\right)_Q=\left(\dfrac{\partial T}{\partial S}\right)_{\Phi}\left(\dfrac{\partial S}{\partial \Phi}\right)_Q+\left(\dfrac{\partial T}{\partial \Phi}\right)_S~
\label{a6}
\end{eqnarray}
taking $T\equiv T(S, \Phi)$.\

\noindent Since the heat capacity diverges at the critical points, hence $\left[\left(\dfrac{\partial T}{\partial S}\right)_{\Phi}\right]_{S=S_i}=0$ and $\left(\dfrac{\partial S}{\partial \Phi}\right)_Q$ are finite at the critical point. Therefore, the above equation becomes
\begin{eqnarray}
\left[\left(\dfrac{\partial T}{\partial \Phi}\right)_Q\right]_{S=S_i}=\left[\left(\dfrac{\partial T}{\partial \Phi}\right)_S\right]_{S=S_i}
\label{a7}
\end{eqnarray}
which implies
\begin{eqnarray}
\left[\left(\dfrac{\partial T}{\partial \Phi}\right)_Q\right]_{r_+=r_{+i}}=\left[\left(\dfrac{\partial T}{\partial \Phi}\right)_{r_+}\right]_{r_+=r_{+i}}~.
\label{a8}
\end{eqnarray}
Therefore, the left hand side of the second Ehrenfest equation is equal to left hand side of the first Ehrenfest equation. Hence, we have
\begin{eqnarray}
\left[\left(\dfrac{\partial \Phi}{\partial T}\right)_Q\right]_{r_+=r_{+i}}=-\dfrac{\omega (r_{+i}^2+2\alpha')r_{+i}^{D-4}}{4Q_i}\left[1+\dfrac{1}{D-3}\sqrt{1+\dfrac{16\pi^2Q_i^2}{b^2\omega^2r_{+i}^{2D-4}}}{}_2F_1\left[\dfrac{D-3}{2D-4},\dfrac{1}{2},\dfrac{3D-7}{2D-4},-\dfrac{16\pi^2Q_i^2}{b^2\omega^2r_{+i}^{2D-4}}\right]\right]~.
\label{a9}
\end{eqnarray}
\noindent Now from eq.(\ref{y}), at the critical points, we have
\begin{eqnarray}
\kappa Q_i=\left[\left(\dfrac{\partial Q}{\partial \Phi}\right)_T\right]_{S=S_i}~.
\label{a10}
\end{eqnarray}
Using the thermodynamic identity $\left(\dfrac{\partial Q}{\partial \phi}\right)_T\left(\dfrac{\partial \Phi}{\partial T}\right)_Q\left(\dfrac{\partial T}{\partial Q}\right)_{\Phi}=-1$ and the definition of $\beta$ in eq.(\ref{y}), we find
\begin{eqnarray}
\kappa Q_i=\left[\left(\dfrac{\partial T}{\partial \Phi}\right)_Q\right]_{S=S_i}Q_i\beta~.
\label{a11}
\end{eqnarray}
\noindent Therefore, the right hand side of the second Ehrenfest equation (\ref{x}) reduces to 
\begin{eqnarray}
\dfrac{\Delta \beta}{\Delta \kappa}=-\left[\left(\dfrac{\partial \Phi}{\partial T}\right)_Q\right]_{S=S_i}~.
\label{a12}
\end{eqnarray}
This can further be written as
\begin{eqnarray}
-\left[\left(\dfrac{\partial \Phi}{\partial T}\right)_Q\right]_{S=S_i}&=&-\left[\left(\dfrac{\partial \Phi}{\partial S}\right)_Q\right]_{S=S_i}\left[\left(\dfrac{\partial S}{\partial T}\right)_Q\right]_{S=S_i}\nonumber\\
&=&-\dfrac{\left[\left(\dfrac{\partial \Phi}{\partial S}\right)_Q\right]_{S=S_i}}{\left[\left(\dfrac{\partial T}{\partial S}\right)_Q\right]_{S=S_i}}~
\label{a13}
\end{eqnarray}
which in turn implies
\begin{eqnarray}
-\left[\left(\dfrac{\partial \Phi}{\partial T}\right)_Q\right]_{S=S_i}\equiv-\left[\left(\dfrac{\partial \Phi}{\partial T}\right)_Q\right]_{r_+=r_{+i}}=-\dfrac{\left[\left(\dfrac{\partial \Phi}{\partial r_+}\right)_Q\right]_{r_+=r_{+i}}}{\left[\left(\dfrac{\partial T}{\partial r_+}\right)_Q\right]_{r_+=r_{+i}}}~.
\label{a14}
\end{eqnarray}
The condition that heat capacity in eq.(\ref{o}) diverges at the critical point gives 
\begin{eqnarray}
\left[\left(\dfrac{\partial \Phi}{\partial Q}\right)_{r_+}\right]_{r_+=r_{+i}}\left[\left(\dfrac{\partial T}{\partial r_+}\right)_Q\right]_{r_+=r_{+i}}=\left[\left(\dfrac{\partial T}{\partial Q}\right)_{r_+}\right]_{r_+=r_{+i}}\left[\left(\dfrac{\partial \Phi}{\partial r_+}\right)_Q\right]_{r_+=r_{+i}}~.
\label{a15}
\end{eqnarray}
Using the above condition and eq.(\ref{a14}) we obtain the right hand side of the second Ehrenfest equation to be
\begin{eqnarray}
\dfrac{\Delta \beta}{\Delta \kappa}=-\dfrac{\omega (r_{+i}^2+2\alpha')r_{+i}^{D-4}}{4Q_i}\left[1+\dfrac{1}{D-3}\sqrt{1+\dfrac{16\pi^2Q_i^2}{b^2\omega^2r_{+i}^{2D-4}}}{}_2F_1\left[\dfrac{D-3}{2D-4},\dfrac{1}{2},\dfrac{3D-7}{2D-4},-\dfrac{16\pi^2Q_i^2}{b^2\omega^2r_{+i}^{2D-4}}\right]\right]~.
\label{a16}
\end{eqnarray}
Eq.(\ref{a9}) and eq.(\ref{a16}) shows the validity of the second Ehrenfest equation at critical points.\

\noindent Finally, we calculate the Prigogine-Defay (PD) ratio \cite{pd} for the system. Using eq.(\ref{w}) and eq.(\ref{a7}), we have
\begin{eqnarray}
\left[\left(\dfrac{\partial \Phi}{\partial T}\right)_Q\right]_{S=S_i}=\left[\left(\dfrac{\partial \Phi}{\partial T}\right)_S\right]_{S=S_i}=-\dfrac{\Delta C_{\Phi}}{T_i Q_i\Delta\beta}~.
\label{a17}
\end{eqnarray}
Now eq.(\ref{a12}) with the above equation gives
\begin{eqnarray}
\Pi=\dfrac{\Delta C_{\Phi}\Delta\kappa}{T_i Q_i(\Delta\beta)^2}=1~.
\label{a18}
\end{eqnarray}
These results show the exact second order nature of phase transition in this  black hole. 
\section{Ruppeiner state space geometry analysis}
\noindent Ruppeiner's state space geometry technique is another approach to study thermodynamic systems. The idea is to write the abstract manifold with the notion of distance from thermodynamic variables and study the curvature in order to study phase transition. This technique has been applied to black hole systems in \cite{rub1}-\cite{rub2}. Definitions of metric coefficients are quite well known \cite{rup}.\
\begin{figure}[b]
		\subfloat[D=5\label{f1}]{{\includegraphics[scale=0.50]{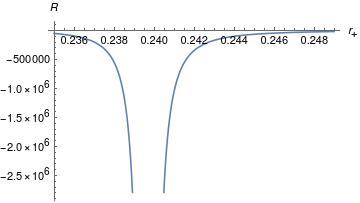}}}
		\qquad
		~~~~~~~~~~
		\subfloat[D=5\label{f2}]{{\includegraphics[scale=0.50]{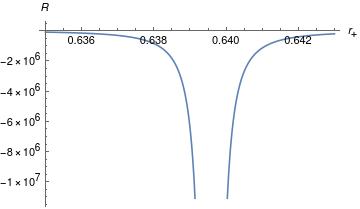}}}
		\qquad
		~~~~~~~~~~
		\subfloat[D=6\label{f3}]{{\includegraphics[scale=0.50]{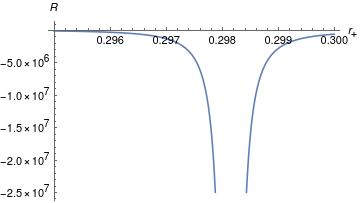}}}
		\qquad
		\subfloat[D=6\label{f4}]{{\includegraphics[scale=0.50]{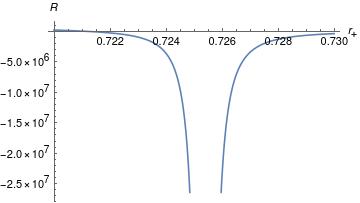}}}
		\qquad
		~~~~~~~~~~
		\subfloat[D=7\label{f5}]{{\includegraphics[scale=0.50]{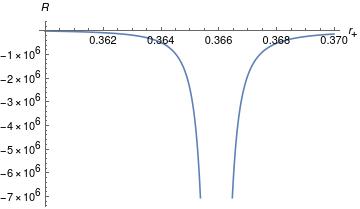}}}
		\qquad
		~~~~~~~~~~
		\subfloat[D=7\label{f6}]{{\includegraphics[scale=0.50]{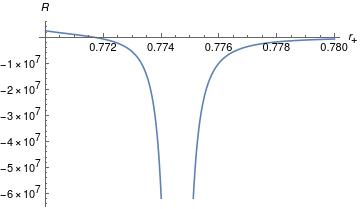}}}
		\qquad		
		\caption{Ruppeiner curvature vs. horizon radius}{(Q=0.13, b=10, $\alpha'$=0.01)}
		\label{rr}
\end{figure}

\noindent The Ruppeiner metric coefficients for the manifold are given by
\begin{eqnarray}
g^{R}_{ij}=-\dfrac{\partial^2 S(x^i)}{\partial x^i \partial x^j}
\label{a19}
\end{eqnarray}
where $x^i=x^i(M, Q)$; $i=1,2$ are extensive variables of the manifold.
\noindent The calculation of the Weinhold metric coefficients is convenient for computational purpose. These are given by
\begin{eqnarray}
g^{W}_{ij}=\dfrac{\partial^2 M(x^i)}{\partial x^i \partial x^j}
\label{a20}
\end{eqnarray}
where $x^i=x^i(S, Q)$; $i=1,2$. It is to be noted that Weinhold geometry is connected to the Ruppeiner geometry through the following relation
\begin{eqnarray}
dS^2_R=\dfrac{dS^2_W}{T}~.
\label{a21}
\end{eqnarray}

\noindent For the black hole spacetime with which we are working, the Ruppeiner metric coefficients can be calculated  to be
\begin{eqnarray}
g_{SS}&=&\dfrac{1}{T}\left[\dfrac{\left(\dfrac{\partial^2M}{\partial r_+^2}\right)_Q}{\left(\dfrac{dS}{dr_+}\right)^2}-\dfrac{\left(\dfrac{\partial M}{\partial r_+}\right)_Q\left(\dfrac{d^2S}{dr_+^2}\right)}{\left(\dfrac{dS}{dr_+}\right)^3}\right]\nonumber\\
g_{SQ}&=&\dfrac{1}{T}\left[\dfrac{\dfrac{\partial^2 M}{\partial r_+\partial Q}}{\dfrac{dS}{dr_+}}\right]\nonumber\\
g_{QQ}&=&\dfrac{1}{T}\left(\dfrac{\partial^2M}{\partial Q^2}\right)_{r_+}~.
\label{a22}
\end{eqnarray}
From eq.(s)(\ref{h}, \ref{j}), we calculate first and second order partial derivatives appearing in eq.(\ref{a22}) to be
\begin{eqnarray}
\dfrac{\partial^2 M}{\partial r_+\partial Q}=-\dfrac{4\pi Q}{\omega r_+^{D-2}}\left(1+\dfrac{16\pi^2Q^2}{b^2\omega^2r_+^{2D-4}}\right)^{-1/2}~
\label{a23}
\end{eqnarray}
\begin{eqnarray}
\dfrac{d^2S}{dr_+^2}=\dfrac{\omega}{4}(D-2)r_+^{D-4}\left((D-3)+\dfrac{2(D-5)\alpha'}{r_+^2}\right)~
\end{eqnarray}
\begin{eqnarray}
\left(\dfrac{\partial M}{\partial r_+}\right)_Q=\dfrac{(D-2)(D-5)\omega\alpha'}{16\pi}r_+^{D-6}+\dfrac{(D-2)(D-3)\omega}{16\pi}r_+^{D-4}+\dfrac{(D-1)(D-2)\omega)}{16\pi}r_+^{D-2}\nonumber\\
+\dfrac{b^2\omega}{4\pi}r_+^{D-2}\left(1-\sqrt{1+\dfrac{16\pi^2Q^2}{b^2\omega^2r_+^{2D-4}}}\right)~
\label{a24}
\end{eqnarray}
\begin{eqnarray}
\left(\dfrac{\partial^2M}{\partial r_+^2}\right)_Q=\dfrac{(D-2)(D-5)(D-6)\omega\alpha'}{16\pi}r_+^{D-7}+\dfrac{(D-2)(D-3)(D-4)\omega}{16\pi}r_+^{D-5}+\dfrac{(D-1)(D-2)^2\omega)}{16\pi}r_+^{D-3}\nonumber\\
+\dfrac{b^2(D-2)\omega}{4\pi}r_+^{D-3}\left(1-\sqrt{1+\dfrac{16\pi^2Q^2}{b^2\omega^2r_+^{2D-4}}}\right)+\dfrac{4(D-2)\pi Q^2}{\omega r_+^{D-1}}\left(1+\dfrac{16\pi^2Q^2}{b^2\omega^2r_+^{2D-4}}\right)^{-1/2}
\label{a25}
\end{eqnarray}
\begin{eqnarray}
\left(\dfrac{\partial^2M}{\partial Q^2}\right)_{r_+}=\dfrac{4\pi}{\omega(D-2) r_+^{D-3}}\left[\left(1+\dfrac{16\pi^2Q^2}{b^2\omega^2r_+^{2D-4}}\right)^{-1/2}+\dfrac{1}{D-3}{}_2F_1\left[\dfrac{D-3}{2D-4},\dfrac{1}{2},\dfrac{3D-7}{2D-4},-\dfrac{16\pi^2Q_i^2}{b^2\omega^2r_{+i}^{2D-4}}\right]\right]~.
\label{a26}
\end{eqnarray}
With these metric coefficients, we can calculate the curvature of the two dimensional manifold, and singularities in the curvature indicate second order phase transition. The formula for the curvature reads 
\begin{eqnarray}
R =-\dfrac{1}{\sqrt{g}}\left[\dfrac{\partial}{\partial r_+}\left(\dfrac{g_{SQ}}{\sqrt{g}g_{SS}}\dfrac{\partial g_{SS}}{\partial Q}-\dfrac{1}{\sqrt{g}}\dfrac{g_{QQ}}{\partial r_+}\left(\dfrac{dr_+}{dS}\right)\right)\left(\dfrac{dr_+}{dS}\right)\right. \nonumber\\
\left.+\dfrac{\partial}{\partial Q}\left(\dfrac{2}{\sqrt{g}}\dfrac{\partial  g_{SQ}}{\partial Q}-\dfrac{1}{\sqrt{g}}\dfrac{\partial g_{SS}}{\partial Q}-\dfrac{g_{SQ}}{\sqrt{g}g_{SS}}\dfrac{\partial g_{SS}}{\partial r_+}\left(\dfrac{dr_+}{dS}\right)\right)\right]~
\label{a27}
\end{eqnarray}
where $g$ is $g_{SS}\times g_{QQ}-g_{SQ}^2$.
We have calculated and plotted $R$ with the horizon radius for fixed values of other parameters in fig. \ref{rr}.\

\noindent These show multiple points of discontinuities, however, all are not physical. It is interesting to note that the Ruppeiner curvature diverges at all the values where the heat capacity is singular. This indicates the second order nature of phase transition in this black hole.

 \section{Conclusion}
\noindent In this work, we have analysed black holes in Gauss-Bonnet gravity with Born-Infeld electrodynamics in AdS background. We calculated the thermodynamic properties associated with the black hole spacetime and calculated the heat capacity. It has been obtained that there are infinite discontinuities in the heat capacity when plotted with the horizon radius with other parameters kept constant. We eastablish that these are the signs of second order phase transitions. First, we compute the free energy of the black hole and observe that there is no cusp in its variation with the horizon radius.
This rules out the possibility of a first order phase transition. The Ehrenfest scheme is then employed to establish the second order nature of the phase transition. We find that the system under consideration obeys both Ehrenfest equations. We then calculate the Prigogine-Defay ratio using Ehrenfest equations which indicate that the divergences are exactly of second order. We also discussed the effects of non-linear electrodynamics on Hawking temperature and heat capacity. We observe that there is a non-trivial effect of the Born-Infeld parameter on the temperature of the black hole. The Hawking temperature  is higher than the black hole with Maxwell term for the same horizon radius and this effect fades away as the horizon radius increases. We also see that the phase transition points shift to lower horizon radius values in the presence of Born-Infeld electrodynamics. The state space geometry technique of Ruppeiner also helped further to reinforce the second order nature of these black hole phase transitions. We calculate the Ruppeiner curvature and again plot it with the horizon radius for the same fixed parameters. Singularities are observed exactly at those points where the heat capacity diverged. Both these methods reconfirm the second order nature of phase transition in these black holes. Reassuringly, we recover the Einstein-Born-Infeld black holes when the Gauss-Bonnet parameter is put to zero. We have also derive a Smarr relation in $D$-spacetime dimensions using scaling arguments and first law of black hole thermodynamics in
which the cosmological constant, Born-Infeld parameters and Gauss-Bonnet parameter are treated as thermodynamic variables.




\def\bibsection{\section*{\refname}}  

\end{document}